\documentclass[12pt, a4paper]{article}
\usepackage{natbib}
\usepackage{graphicx}
\usepackage{amsmath}

\title{A Study of Error Variance Estimation in Lasso Regression}
\author{Stephen Reid$^1$, Robert Tibshirani$^2$ and Jerome Friedman$^1$}
\date{\normalsize $^1$ Department of Statistics, Stanford University, $^2$ Departments of Health, Research \& Policy, and Statistics, Stanford University}

\newtheorem{theorem}{Theorem}[section]
\newtheorem{lemma}[theorem]{Lemma}

\begin{document}

\maketitle

\begin{abstract}
 Variance estimation in the linear model when $p > n$ is a difficult problem. Standard least squares estimation techniques do not apply. Several variance estimators have been proposed in the literature, all with accompanying asymptotic results proving consistency and asymptotic normality under a variety of assumptions. 

It is found, however, that most of these estimators suffer large biases in finite samples when true underlying signals become less sparse with larger per element signal strength. One estimator seems to be largely neglected in the literature: a residual sum of squares based estimator using Lasso coefficients with regularisation parameter selected adaptively (via cross-validation). 

In this paper, we review several variance estimators and perform a reasonably extensive simulation study in an attempt to compare their finite sample performance. It would seem from the results that variance estimators with adaptively chosen regularisation parameters perform admirably over a broad range of sparsity and signal strength settings. Finally, some intial theoretical analyses pertaining to these types of estimators are proposed and developed.

\end{abstract}

\section{Introduction}

Consider the linear model

\[
Y = X\beta + \epsilon
\]
where $Y$ is an $n$-vector of independently distributed responses, $X$ an $n \times p$ matrix with individual specific covariate vectors as its rows and $\epsilon$ an $n$-vector of i.i.d random variables (usually assumed Gaussian) each with mean $0$ and variance $\sigma^2$.

When $p > n$, one cannot estimate the unknown coefficient vector $\beta$ uniquely via standard least squares methodology. In fact, it is probably ill-advised to use least squares to estimate the vector even when $p \leq n$ and $p$ close to $n$, since standard errors are likely to be high and parameter estimates unstable. In this instance, if one can assume that $\beta$ is reasonably sparse with many zero entries, a successful method for selecting the nonzero elements of $\beta$ and estimating them is the Lasso estimator proposed by \citet{lasso}, obtained by minimising

\[
 \frac{1}{2}||Y - X\beta||_2^2 + \lambda||\beta||_1 
\]
where the parameter $\lambda$ is predetermined and controls the amount of regularisation. The higher the value of $\lambda$, the more elements of the estimated $\beta$ vector are set to $0$ and the more the nonzero entries are shrunken toward $0$. Smaller $\lambda$ implies less regularisation and more nonzero $\beta$ with larger (absolute) coefficients.

Much has been written about the model selection and prediction properties of this class of estimators, but it is only recently that people have turned to developing significance tests for the estimated coefficients. Examples include \citet{LockhartEtAl2012} and \citet{JavanMonta2013}. Each of these requires an estimate of error variance $\sigma^2$ to plug into their chosen test statistics. A good estimate of $\sigma^2$ is required. The problem of estimating error variance when $p > n$ is interesting in its own right and several estimators have been proposed by different authors. 

The aim of this paper is to review some of these estimators and to run an extensive simulation experiment comparing their estimation performance over a broad range of parameter vector sparsity and signal strength settings. Perhaps an unbiased comparison of these estimators may reveal the most promising estimator, helping to guide research into fruitful directions. In particular, a promising estimator seems to be

\[
\hat{\sigma}^2 = \frac{1}{n - \hat{s}_{\hat{\lambda}}}||Y - X\hat{\beta}_{\hat{\lambda}}||_2^2
\]
where $\hat{\beta}_\lambda$ is the Lasso estimate at regularisation parameter $\lambda$, $\hat{\lambda}$ is selected via cross-validation and $\hat{s}_{\hat{\lambda}}$ is the number of nonzero elements in $\hat{\beta}_{\hat{\lambda}}$

\section{Review of error variance estimators}

In this section, we review some of the error variance estimators proposed recently and list some of their theoretical properties, as well as the assumptions under which these properties hold.

\subsection{The oracle}
The ideal variance estimator is the oracle estimator:

\begin{equation}
 \hat{\sigma}_O^2 = \frac{1}{n}\sum_{i = 1}^n(Y_i - X_i^\prime\beta^*)^2
 \label{sigEstiOracle}
\end{equation}
where $\beta^*$ is the true (unknown) coefficient vector with $s$ nonzero elements. This estimator (times $n$) has a $\chi^2$ distribution with $n$ degrees of freedom and serves as a sample variance for the zero mean $\epsilon$. Obviously this is not a viable estimator in practice, because we do not know $\beta^*$. However, it is useful for comparison purposes in a simulation study.

\subsection{Residual sum of squares based estimators} 
\citet{FanVar} consider estimators of the form:

\begin{equation}
 \hat{\sigma}_{L, \lambda_n}^2 = \frac{1}{n - \hat{s}_{L, \lambda_n}}\sum_{i = 1}^n(Y_i - X_i^\prime\hat{\beta}_{\lambda_n})^2
\end{equation}
where $\beta_\lambda$ is the Lasso coefficient vector estimate, and $\hat{s}_{L, \lambda}$ the number of nonzero elements of this vector, at regularisation parameter $\lambda$. \citet{GreenRit2004} show estimators of this form to be consistent for $\sigma^2$ under some technical conditions on the population moments of $Y$ and $X$. Consistency holds if $\lambda_n = O(\sqrt{\frac{log(p)}{n}})$.

\citet{FanVar} show that this estimator has a limiting zero mean normal distribution as $n \rightarrow \infty$, $\frac{s\log(p)}{\sqrt{n}} \rightarrow 0$ and $\lambda_n \propto \sigma\sqrt{\frac{log(p)}{n}}$. Furthermore, this limiting distribution has the same variance as the asymptotic variance of the oracle estimator.

Their results are gleaned by making assumptions on the elements of matrix $X$ (assumed to be bounded absolutely) and the so-called \textit{sparse-eigenvalues}. The smallest and largest sparse eigenvalues are defined respectively as:
\[
 \phi_{min}(m) = \min_{M: |M| \leq m}\lambda_{min}(\frac{1}{n}X_M^TX_M)
\]
and
\[
 \phi_{max}(m) = \max_{M: |M| \leq m}\lambda_{max}(\frac{1}{n}X_M^TX_M)
\]
where $M$ is a set of integers selected from $\{1, 2, ..., p\}$, $X_M$ is the $n \times M$ matrix obtained by selecting columns from $X$ indexed by elements of $M$ and $\lambda_{min}(A)$ and $\lambda_{max}(A)$ are the smallest and largest eigenvalues of matrix $A$ respectively. Assumptions are made bounding the asymptotic behaviour of these sparse eigenvalues. A lower bound on the smallest sparse eigenvalue seems to be particularly important. These types of assumptions seem to be quite prevalent in the literature that pertains to our problem.

Although heartening, results of this kind are not useful in practice. The choice of $\lambda$ is very important in the pursuit of an accurate finite sample estimator. Its size controls both the number of variables selected and the degree to which their estimated coeffcients are shrunk to zero. Set $\lambda$ too large and we do not select all signal variables, leading to rapidly degrading performance (exhibited mostly by large upward bias) when the true $\beta$ becomes less sparse with larger signal per element. On the other hand, should we set $\lambda$ too small, we would select many noise variables, allowing spurious correlation to decrease our variance estimate, leading to substantial downward bias. Simulation results seem to suggest there is a fine balance to be maintained when selecting the appropriate $\lambda$.

\subsection{Cross-validation based estimators}

Considerations around the selection of an appropriate $\lambda$ lead us inexorably toward an adaptive selection method. In particular, one can define:

\begin{equation}
 \hat{\sigma}_{L, \hat{\lambda}}^2 = \frac{1}{n - \hat{s}_{L, \hat{\lambda}}}\sum_{i = 1}^n(Y_i - X_i^\prime\hat{\beta}_{\hat{\lambda}})^2
 \label{sigEstiCVLasso}
\end{equation}
where $\hat{\lambda}$ is selected using $K$-fold cross-validation. $K$ is usually set to $5$ or $10$. Our simulation results suggest that this estimator is robust to changes in signal sparsity and strength, more so than its competitors. \citet{FanVar} lament the downward bias of this estimator. They claim that it is affected by spurious correlation. Although this downward bias seems to be borne out in our simulation results, it does not seem too large and stems from a heavy left tail in its empirical distribution. The median estimate tends to be very close to the true $\sigma^2$ under a surprisingly broad range of sparsity and signal strength settings.

Sadly, very little theory exists detailing the properties of this estimator. \citet{Homrig2013} prove a result on the persistence of this estimator that can, with a suitable sparsity assumption on the true $\beta$, be adapted to a consistency result for an estimate closely resembling $\hat{\sigma}^2_{L, \hat{\lambda}}$. 

An implication of their result is that if the true underlying coefficient vector $\beta^*$ is sufficiently sparse, i.e. $||\beta^*||_1 = o\left(\left(\frac{n}{\log(n)}\right)^{\frac{1}{4}}\right)$, then
\[
\frac{n - \hat{s}}{n}\hat{\sigma}^2_{L, \hat{\lambda}} \stackrel{P}{\rightarrow} \sigma^2
\]
If one can assume, as do \citet{FanVar}, that $\hat{s} = o_P(n)$, then $\hat{\sigma}^2_{L, \hat{\lambda}}$ is also consistent. We are not aware of a proof of this for cross-validation though. Nothing is said about the finite sample distribution of this estimator, or whether any asymptotic distribution obtains for that matter.

\citet{FanVar} propose two other cross-validation based variance estimators. The first defines the $K$ cross-validation folds as $\{D_1, D_2, ..., D_K\}$ and computes:

\begin{equation}
 \hat{\sigma}_{CVL}^2 = \min_{\lambda}\frac{1}{n}\sum_{k = 1}^K\sum_{i \in D_k}(Y_i - X_i^\prime\hat{\beta}^{(-k)}_\lambda)^2
 \label{sigEstiCVLasso2}
\end{equation}
where $\hat{\beta}^{(-k)}_\lambda$ is the Lasso estimate at $\lambda$ over the data after the $k^{th}$ fold is omitted. They try $K = 5, 10,$ and $n$, the latter corresponding to leave-one-out cross-validation. They find in their simulations that the estimate is consistently above the true error variance. We found the same tendency and this estimator is omitted from the simulation study exposition in the next section.

A second estimator uses cross-validation to select the optimal regularisation parameter $\hat{\lambda}$ and then finds the set of indices corresponding to nonzero entries in $\hat{\beta}_{\hat{\lambda}}$. Call this set $\hat{M}$. The ``na\"{i}ve'' two-stage Lasso estimator is then defined as:

\begin{equation}
 \hat{\sigma}_{NL}^2 = \frac{1}{n - |\hat{M}|}||(I - X_{\hat{M}}(X_{\hat{M}}^\prime X_{\hat{M}})^{-1}X_{\hat{M}}^\prime)Y||^2_2
 \label{sigEstiLassoNaive}
\end{equation}
This estimator suffers from downward bias for sparse $\beta$, because the Lasso tends to overselect (including the vast majority of signal variables and a few noise variables). Least squares estimates of parameters are not shrunk toward zero and inclusion of additional noise variables (that seem well correlated with the response) drives down the variance estimate. \citet{WasserRoeder2009} demonstrate the overselection property of the Lasso. The downward bias of this estimator is made apparent in our simulation results.

\subsection{Refitted Cross-Validation (RCV) estimator}

In an attempt to overcome the downward bias caused by spurious correlation in the na\"{i}ve Lasso estimator, \citet{FanVar} propose a refitted cross-validation (RCV) estimator. They split the dataset into two (roughly) equal parts $X^{(1)}$ and $X^{(2)}$. On the first part, $X^{(1)}$, they fit the Lasso, using cross-validation to determine the optimal regularisation parameter $\hat{\lambda}_1$ and corresponding set of nonzero indices $\hat{M}_1$. Using those columns in $X^{(2)}$ indexed by $\hat{M}_1$ they obtain the following variance estimate:
\[
\hat{\sigma}_1^2 = \frac{1}{n - |\hat{M}_1|}||(I - X_{\hat{M}_1}^{(2)}(X_{\hat{M}_1}^{(2)\prime} X_{\hat{M}_1}^{(2)})^{-1}X_{\hat{M}_1}^{(2)\prime})Y||^2_2
\]
They then repeat the mirror image procedure on $X^{(2)}$, obtaining $\hat{\lambda}_2$, $\hat{M}_2$ and $\hat{\sigma}_2^2$. The RCV variance estimate is the obtained as:

\begin{equation}
 \hat{\sigma}^2_{RCV} = \frac{\hat{\sigma}^2_1 + \hat{\sigma}^2_2}{2}
 \label{sigEstiRCV}
\end{equation}
The authors prove consistency and asymptotic normality (with asymptotic variance the same as that of the oracle estimator) of this estimator under slightly weaker conditions used for proving similar results for $\hat{\sigma}_{L, \lambda_n}^2$. They argue that breaking up the dataset counters the effect of spurious correlation, since spurious noise variables selected on one half are unlikely to produce significant least squares parameter estimates on the second half, reducing the negative bias associated with the overselection of the Lasso selector. 

Theoretical results aside, the finite sample performance of this estimator seems to suffer when $\beta$ is less sparse and has larger signal per element. The plug in Lasso estimator $\hat{\sigma}^2_{L, \hat{\lambda}}$ remains anchored around the true $\sigma^2$ for a broader array of sparsity and signal strength settings.

\subsection{SCAD estimator}
The Lasso is just one method for selecting the variables to have nonzero coefficients in our variance estimator. Any other valid variable selection method could be used to estimate error variance in the spirit of $\hat{\sigma}^2_{L, \hat{\lambda}}$. One such method is the Smoothly Clipped Absolute Deviation Penalty (SCAD) of \citet{FanSCAD}. Instead of using an $\ell_1$ penalty, they minimise:

\[
 \frac{1}{2}||Y - X\beta||_2^2 + \sum_{j = 1}^pp_\lambda(|\beta_j|)
\]
where $p^\prime_\lambda(\theta) = \lambda\left(I(\theta \leq \lambda) + \frac{(a\lambda - \theta)_+}{(a - 1)\lambda}I(\theta > \lambda)\right)$ for some $a > 2$ (usually $3.7$) and $\theta > 0$. This penalty is chosen for its good model selection properties. Although no longer a convex criterion, the authors claim to have a stable and reliable algorithm for determining the optimal $\beta$ with good properties. Indeed, their simulations seem to suggest that SCAD outperforms the Lasso at variable selection in the low noise case when both have their regularisation parameters chosen by cross-validation.

Given a method with good variable selection performance (i.e. it selects the signal variables and few or none of the noise variables), we have a hope of mimicking an oracle estimator that is privy to the correct $\beta$. \citet{FanVar} define their SCAD variance estimator as:

\begin{equation}
\hat{\sigma}^2_{SCAD} = \frac{1}{n - \hat{s}_{\hat{\lambda}}}||Y - X\hat{\beta}_{SCAD, \hat{\lambda}}||_2^2
\label{sigEstiSCAD}
\end{equation}
where $\hat{\beta}_{SCAD, \hat{\lambda}}$ is the SCAD estimate of $\beta$ at the regularisation parameter $\hat{\lambda}$ selected by cross-validation. Again, consistency and asymptotic normality can be shown for this estimator with an appropriately chosen, deterministic regularisation parameter sequence $\lambda_n$. Our simulations suggest that it performs comparably to $\hat{\sigma}^2_{L, \hat{\lambda}}$.

\subsection{Scaled Sparse Linear Regression estimators}
\citet{StadlerBuhlmann2010} introduce the notion of estimating jointly the parameter vector and error variance in the context of mixture regression models. \citet{SZComment} refine this notion for the non-mixture case and explore the properties of this new estimator in \citet{SZBiomet}.

In particular, the latter pair proposes the joint optimisation in $(\beta, \sigma)$ of the jointly convex criterion (called the ``scaled Lasso'' criterion):

\begin{equation}
\frac{||Y - X\beta||^2_2}{2n\sigma} + \frac{\sigma}{2} + \lambda_0||\beta||_1
\label{sigEstiScaledLasso}
\end{equation}
where $\lambda_0$ is some predetermined fixed parameter. An iterative, alternating optimisation algorithm is given where, given a current estimate $\hat{\beta}^{current}$, parameter estimates are updated as:
\[
\hat{\sigma} = \frac{||Y-X\hat{\beta}^{current}||_2}{\sqrt{n}}
\]
\[
\lambda = \hat{\sigma}\lambda_0
\]
\[
\hat{\beta}^{current} = \hat{\beta}_{\lambda}
\]
where $\hat{\beta}_\lambda$ is the Lasso estimate of $\beta$ at regularisation parameter $\lambda$. These steps are interated until the parameter estimates converge.

The authors go on to show consistency, asymptotic normality and oracle inequalities for this estimator under a \textit{compatibility} assumption (detailed in their paper) and assumptions on the sparse eigenvalues of $X$. The finite sample success of this method, however, hinges on the choice of $\lambda_0$. The asymptotic results hold when $\lambda_0 \propto \sqrt{\frac{\log(p)}{n}}$, but finite sample accuracy will depend greatly on an appropriate choice of the proportionality constant. Simulation results from their paper suggest that $\sqrt{2}$ is a good choice for the proportionality constant, but our simulation results show rapid degradation as the true $\beta$ becomes less sparse with larger per element signal.

Another estimator proposed by \citet{SZBiomet} uses the scaled Lasso criterion to find $\hat{M}_{SZ}$ - the set of indices corresponding to nonzero $\hat{\beta}^{current}$ after the final iteration. Once obtained, another estimator is defined as:

\begin{equation}
\hat{\sigma}_{SZLS}^2 = \frac{1}{n - |\hat{M}_{SZ}|}||(I - X_{\hat{M}_{SZ}}(X_{\hat{M}_{SZ}}^{\prime} X_{\hat{M}_{SZ}})^{-1}X_{\hat{M}_{SZ}}^{\prime})Y||^2_2
\label{sigEstiScaledLassoLS}
\end{equation}
The authors tout the finite sample accuracy of this estimator. 

In a recent paper, \citet{SZ2013} propose a different value for $\lambda_0$. With this value, tighter error bounds are achieved than in their previous paper. In particular, they propose
\begin{equation}
 \lambda_0 = \sqrt{2}L_n(\frac{k}{p})
 \label{SZRegParam}
\end{equation}
with $L_n(t) = \frac{1}{\sqrt{n}}\Phi^{-1}(1-t)$, where $\Phi$ is the standard Gaussian cdf and $k$ is the solution to
\[
k = L_1^4(\frac{k}{p}) + 2L_1^2(\frac{k}{p})
\]

A least squares after scaled Lasso estimator (as in \eqref{sigEstiScaledLassoLS}) is also proposed for this level of the regularisation parameter. All four of the scaled Lasso estimators were included in our simulation study.

\subsection{Method of Moments estimators}
\citet{Dicker2014} takes a different tack. Instead of attempting to emulate the sum of squares estimator of standard least squares regression methodology, he makes distributional assumptions on both the errors $\epsilon$ and the columns of the predictor matrix $X$.

He retains the standard assumption that $\epsilon \sim N_n(0, \sigma^2I_n)$, although he makes it at the outset; the subsquent derivation of his estimator depending heavily on this assumption. Furthermore, he assumes that each of the $n$ rows of $X$ (call the $i^{th}$ one $x_i$) is normally distributed: $x_i \sim N_p(0, \Sigma)$. Also, all $\epsilon_i$ and $x_i$ are assumed independent.

These distributional assumptions allow one to compute the expectations of the quantities $||y||^2$ and $||X^\prime y||^2$. Equating these moments to their sample counterparts enables one to derive estimators for $\sigma^2$. 

He proposes two estimators. The first holds when we assume $\Sigma = I_p$:

\begin{equation}
\hat{\sigma}^2_{D1} = \frac{p + n + 1}{n(n+1)}||y||^2 - \frac{1}{n(n+1)}||X^\prime y||^2 
\label{sigEstiDickerI}
\end{equation}
while a second estimator is an approximate method of moments estimator for the case of general $\Sigma$:

\begin{equation}
\hat{\sigma}^2_{D2} = \left[1 + \frac{p\hat{m}_1^2}{(n+1)\hat{m}_2}\right]\frac{1}{n}||y||^2 - \frac{\hat{m}_1}{n(n+1)\hat{m}_2}||X^\prime y||^2
\label{sigEstiDickerSigma}
\end{equation}
where

\[
\hat{m}_1 = \frac{1}{p}{\rm tr}\left(\frac{1}{n}X^\prime X\right), \hat{m}_2 = \frac{1}{p}{\rm tr}\left[\left(\frac{1}{n}X^\prime X\right)^2\right] - \frac{1}{pn}\left[{\rm tr}\left(\frac{1}{n}X^\prime X\right)\right]^2
\]

He shows how these estimators are consistent for $\sigma^2$ and have asymptotic Gaussian distributions.

\section{A simulation study}
The merits of each of the estimators above are demonstrated by the authors who devised them. Asymptotic results are gleaned for each and simulation studies run to show some real world applicability. In this section we exact upon the entire collection a fairly extensive simulation study. In the study we control the sparsity of the underlying true $\beta$ vector as well as its signal-to-noise ratio (SNR). The correlation between columns of the $X$ matrix is also controlled. The aim of the study is to reveal the strengths and weaknesses of the estimators (and the sparsity-signal strength combinations in which these are most clearly revealed). In particular, we would like to ascertain which estimator provides reasonable estimates of the error variance over the broadest range of sparsity and signal strength settings. 

Use of the Lasso (and other sparsity inducing coefficient estimators) makes a large bet on sparsity. Most of the good results obtained for this class of estimators make some crucial assumptions about the sparsity of the underlying ground truth. The variance estimators above also rely heavily on the notion of finding \textit{the} small set of nonzero coefficients and using them to remove the signal from the response, leaving only random error, the variance of which can then be obtained. In practice though, we are rarely completely certain about the extent of sparsity of the ground truth. A variance estimator that performs reasonably over a broad range of ground truth settings lends some peace of mind.

\subsection{Simulation parameters}

All simulations are run at a sample size of $n = 100$. Four different values for the number of total predictors are considered: $p = 100, 200, 500, 1000$. Elements of the predictor matrix $X$ are generated randomly as $X_{ij} \sim N(0,1)$. Correlation between columns of $X$ is set to $\rho$.

The true $\beta$ is generated in steps. First, the number of nonzero elements is set to $p_{nz} = \lceil n^\alpha \rceil$. The parameter $\alpha$ controls the degree of sparsity of $\beta$: the higher the $\alpha$; the \textit{less} sparse the $\beta$. It ranges between $0$ and $1$, except when we set it to $-\infty$ to enforce $\beta = 0$. The indices corresponding to nonzero $\beta$ are then selected randomly. Their values are set equal to that of a random sample from a $Laplace(1)$ distibution. The elements of the resulting $\beta$ are then scaled such that the signal-to-noise ratio, defined as $\frac{\beta^\prime\Sigma\beta}{\sigma^2}$, is some predetermined value, $snr$. Here $\Sigma$ is the covariance matrix of the elements of a single row of $X$.

Simulations were run over a grid of values for each of the parameters described above. In particular, 

\begin{itemize}
  \item $\rho = 0, 0.2, 0.4, 0.6, 0.8$.
  \item $\alpha = 0.1, 0.3, 0.5, 0.7, 0.9$.
  \item $snr = 0.5, 1, 2, 5, 10, 20$.
\end{itemize}

At each setting of the parameters, $B = 100$ replications of each of a collection of error variance estimators were obtained. The collection of estimators considered comprises of:

\begin{itemize}
  \item The oracle estimator of Equation~\eqref{sigEstiOracle}.
  \item The cross-validation based Lasso estimator $\hat{\sigma}_{L, \hat{\lambda}}^2$ of Equation~\eqref{sigEstiCVLasso}, denoted \textit{CV\_L} in the simulation output.
  \item The na\"{i}ve Lasso estimator $\hat{\sigma}^2_{NL}$ in Equation~\eqref{sigEstiLassoNaive}, denoted \textit{CV\_LS}.
  \item The SCAD estimator $\hat{\sigma}^2_{SCAD}$ in Equation~\eqref{sigEstiSCAD}.
  \item The RCV estimator $\hat{\sigma}^2_{RCV}$ in Equation~\eqref{sigEstiRCV}.
  \item The scaled Lasso estimator of \citet{SZBiomet} in Equation~\eqref{sigEstiScaledLasso}, denoted \textit{SZ} in output.
  \item Its least-squares-after-scaled-Lasso version in Equation~\eqref{sigEstiScaledLassoLS}, denoted \textit{SZ\_LS}.
  \item The scaled Lasso estimator of \citet{SZ2013} with smaller regularisation parameter of Equation~\eqref{SZRegParam}, denoted \textit{SZ2}.
  \item Its least-squares-after-scaled-Lasso version, denoted \textit{SZ2\_LS}.
  \item Two method of moments estimators of \citet{Dicker2014} from Equations~\eqref{sigEstiDickerI} and~\eqref{sigEstiDickerSigma}, denoted \textit{D1} and \textit{D2} respectively.
\end{itemize}

All figures, tables and results are quoted in terms of the standard deviation estimate. Results were obtained for true error variance values $\sigma = 1, 3$.

\subsection{No signal case: $\beta = 0$}
\begin{figure}[htb]
 \centering
 \includegraphics[width=150mm, height=110mm]{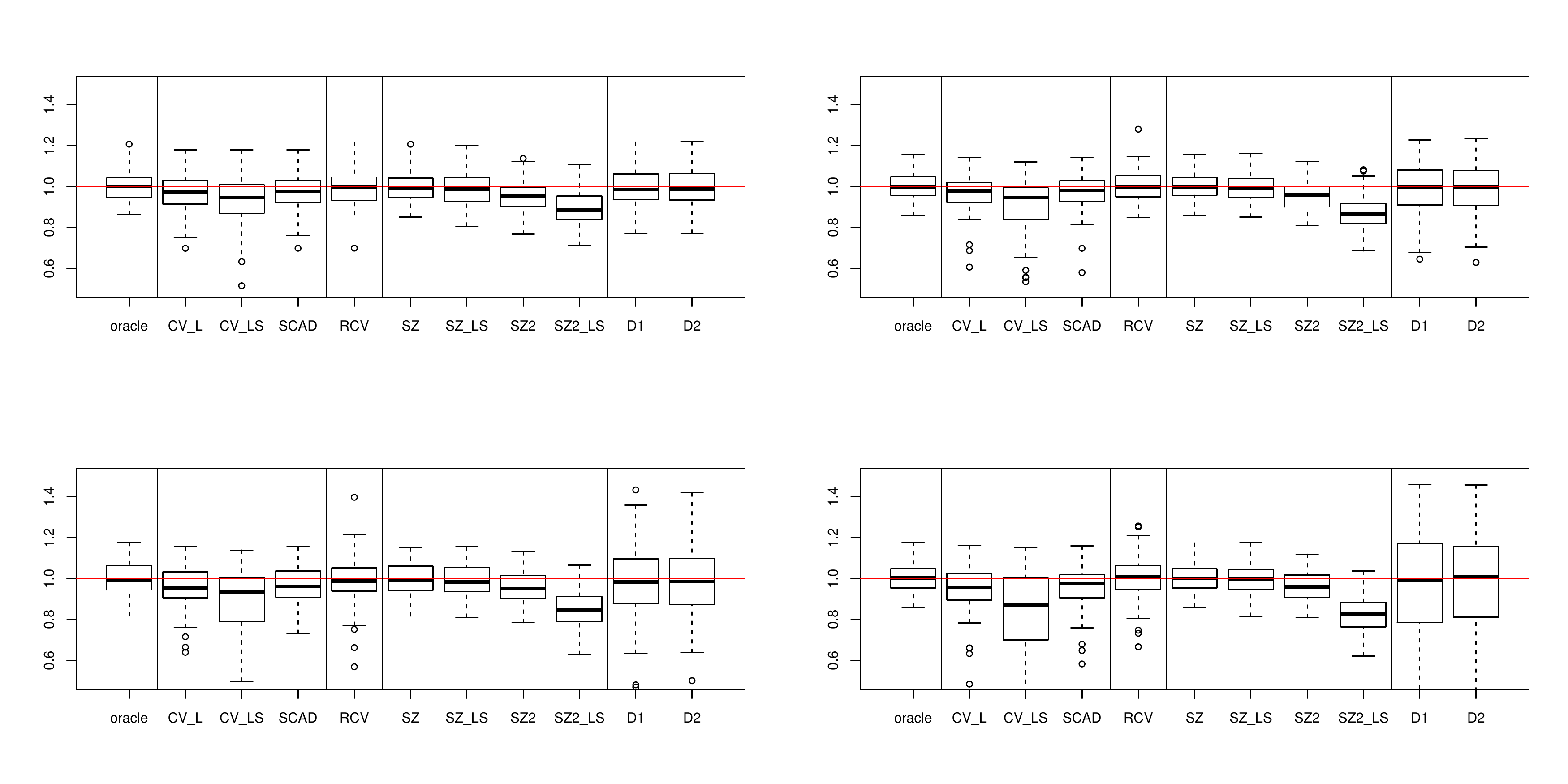}
 \caption{\emph{Standard deviation estimates for $\beta = 0$ case. Sample size $n = 100$, predictors $p = 100, 200, 500, 1000$ moving left to right along rows. $\rho = 0$.}}
 \label{figRho0Beta0}
\end{figure}

We first consider the edge case where $\alpha = -\infty$, forcing $\beta = 0$. Obviously the $snr$ is irrelevant here, because we have no signal. Figure~\ref{figRho0Beta0} shows boxplots of the replications of the standard deviation estimates when $\rho = 0$ and $\sigma = 1$. True $\sigma$ is indicated by the horizontal red line, for reference.

It is apparent that the CV\_L, SCAD and SZ2 estimators are slightly downward biased, whereas the RCV, SZ and SZ\_LS estimators seem to be unbiased. The least-squares-after-Lasso-CV estimator (CV\_LS) is considerably downward biased  (as in SZ2\_LS). This probably stems from the tendency of Lasso to overselect when the regularisation parameter is chosen via CV. The relatively large set of predictors chosen, coupled with the ill effects of spurious correlation when estimating via least squares, probably contribute most significantly to this downward bias.

The method of moments estimators tend to be median unbiased, but their variances increase considerably as $p$ increases relative to $n$. This increase in variance is most pronounced in the bottom right panel ($p = 1000$), where the method of moments estimators have significantly larger variance than the rest.

Median biases for each of the estimators are tabulated in Table~\ref{tabRho0Beta0} for the different \textit{n-p} combinations. This is defined as ${\rm{median}}_{b = 1,2,...,B}\{\hat{\sigma}_b\} - \sigma$, where $\hat{\sigma}_b$ is the $b^{th}$ replication of the standard deviation estimate of interest. It would seem that median biases for CV\_L, CV\_LS and SCAD increase (absolutely) as $p$ increases. Although lamentable, the biases of CV\_L and SCAD are not that large, particularly when compared to the biases of the other estimators when we start increasing the signal (see below). Furthermore, in practice, the assumption is often that there is indeed a signal. This is usually the point of the real world study. This particular setup then, may not be encountered too often in practice.

\begin{table}[htb]
  \centering
  \begin{tabular} {l r r r r}
    \hline
    \hline
    Method & p = 100 & p = 200 & p = 500 & p = 1000 \\ [0.5ex]
    \hline
    Oracle & 0.0009 & -0.0019 & -0.0071 & 0.0041 \\
    CV\_L & -0.0251 & -0.0206 & -0.0439 & -0.0415 \\
    CV\_LS & -0.0519 & -0.0527 & -0.0641 & -0.1302 \\
    SCAD & -0.0232 & -0.0184 & -0.0386 & -0.0227\\
    RCV & -0.0004 & -0.0030 & -0.0110 & 0.0112 \\
    SZ & -0.0046 & -0.0027 & -0.0075 & 0.0013 \\
    SZ\_LS & -0.0118 & -0.0071 & -0.0160 & -0.0001\\
    SZ2 & -0.0451 & -0.0408 & -0.0485 & -0.0397\\
    SZ2\_LS & -0.1139 & -0.1341 & -0.1518 & -0.1739 \\
    D1 & -0.0140 & -0.0016 & -0.0167 & -0.0042 \\
    D2 & -0.0111 & -0.0028 & -0.0143 & 0.0091\\
    \hline
    
  \end{tabular} 
  \caption{\emph{Median biases of standard deviation estimators. No signal, $\sigma = 1$, $\rho = 0$}.}
  \label{tabRho0Beta0}
\end{table}

We also notice from Figure~\ref{figRho0Beta0} the tight clustering of the estimates around the true $\sigma$. None of the clusterings are as tight as that of the oracle, but on the whole, all the standard deviation estimates seem to have low variance (except for CV\_LS, D1 and D2). CV\_L and RCV seem to produce rare outlier estimates, with those from CV\_L always coming in below the true $\sigma$. The distribution of CV\_L seems to be skewed to the left, which may make it difficult to analyse, particularly when one wants to ascertain the distribution of a test statistic using this variance estimator. The effort may be merited though, as we will see below that this estimator performs admirably over a broad range of sparsity and signal strength assumptions.

Correlation between the columns of the predictor matrix seems to have little effect on the performance of each of our estimators. Curves (not shown) depicting median standard deviation estimates as a function of predictor correlation $\rho$ all seem relatively flat, with the estimators retaining their properties as discussed above. Similar looking curves are obtained for the high noise case, $\sigma = 3$ (also not shown).

\subsection{Effect of sparsity: changing $\alpha$}

The true value of the CV\_L estimator becomes apparent once we consider different sparsity levels and signal strength settings. It should be noted that each of the estimators eventually breaks down when signals become non-sparse and large. This is reflected by the very large upward biases in all of the estimators. The question then is not whether we can find a silver bullet for all conceivable ground truths, but rather one that performs reasonably for a broad range of possible ground truths.

Our first consideration in the quest for such a broadly applicable estimator is the effect of decreased sparsity. In our simulation, the sparsity level is controlled by changing the value of $\alpha$. The higher the $\alpha$; the less sparse the ground truth $\beta$ becomes. 

\begin{figure}[htb]
  \centering
  \includegraphics[width=140mm, height=100mm]{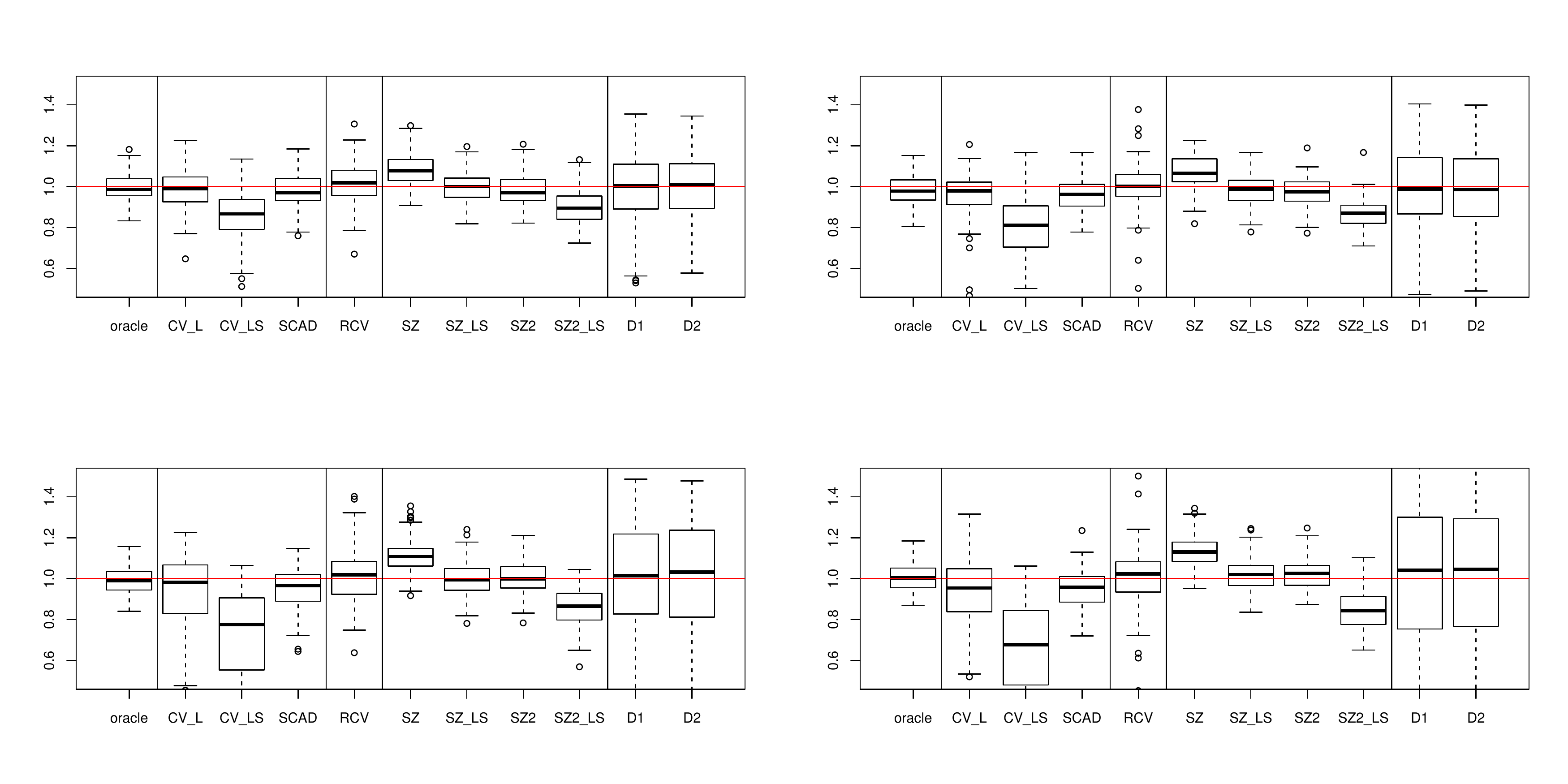}
  \caption{\emph{Standard deviation estimates for $\alpha = 0.1$ (sparse). Sample size $n = 100$, predictors $p = 100, 200, 500, 1000$ moving left to right along rows. $\rho = 0$, $snr = 1$.}}
  \label{figAlpha0.1Rho0SNR1}
\end{figure}

Figure~\ref{figAlpha0.1Rho0SNR1} shows the boxplots of standard deviation estimates when $\alpha = 0.1$ and $snr = 1$ in the uncorrelated case ($\rho = 0$). Notice that the median bias of the CV\_L estimator seems to have decreased, while that of the SCAD estimator has remained negative, roughly of the same size as the no signal case. Downward bias in CV\_LS now seems more pronounced, while the SZ estimator has become upwardly biased, with bias increasing with $p$. SZ\_LS performs best, being unbiased with a tight distribution around its median. CV\_L and RCV perform comparably.

\begin{figure}[htb]
  \centering
  \includegraphics[width=140mm, height=100mm]{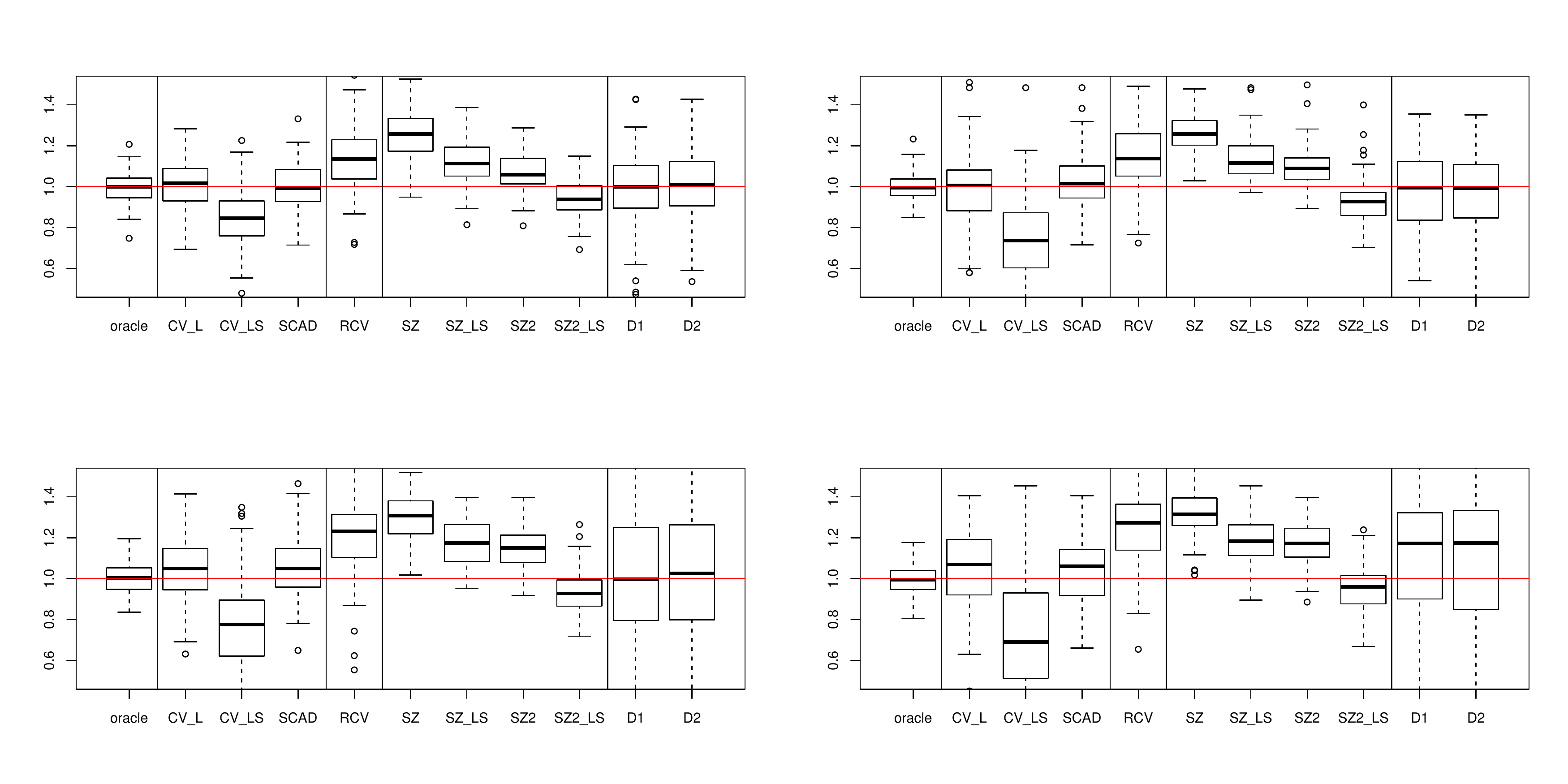}
  \caption{\emph{Standard deviation estimates for $\alpha = 0.5$ (less sparse). Sample size $n = 100$, predictors $p = 100, 200, 500, 1000$ moving left to right along rows. $\rho = 0$, $snr = 1$.}}
  \label{figAlpha0.5Rho0SNR1}
\end{figure}

The narrative changes quite dramatically when we set $\alpha = 0.5$, as in Figure~\ref{figAlpha0.5Rho0SNR1}. Here we see that the CV\_L and SCAD estimators are the only two estimators without substantial biases in either direction. RCV, SZ and SZ\_LS have all become biased upward by roughly $20\%$, while SZ2 becomes increasingly more upwardly biased as $p$ increases, starting with a bias of about $6\%$ at $p = 100$, growing to about $17\%$ when $p = 1000$.

\subsubsection{Explaining the biases}
In attempt to understand why these biases obtain, consider the oracle estimator:
\[
\hat{\sigma}^2_O = \frac{1}{n}\sum_{i = 1}^n(Y_i - X_i^\prime\beta^*)^2
\]

The success of this estimator hinges on the fact that it knows the true $\beta$ (which we call $\beta^*$). It is able to remove all the signal from the observed $Y_i$, leaving only the errors $\epsilon_i$, the variance of which we wish to measure. 

Other estimators (except the method of moment estimators) attempt to emulate the form of the oracle, but none of them are privy even to the set of non-zero $\beta_j$, let alone their true values. Each of these estimators needs to estimate the set of non-zero estimators and then place values on their coefficients. Departures from oracle performance occur when true signal variables are not selected (false negatives), irrelevant variables are selected (false positives) and when estimates of the coefficient values do not match their true underlying values.

\begin{figure}[htb]
\centering
\includegraphics[width=140mm, height=100mm]{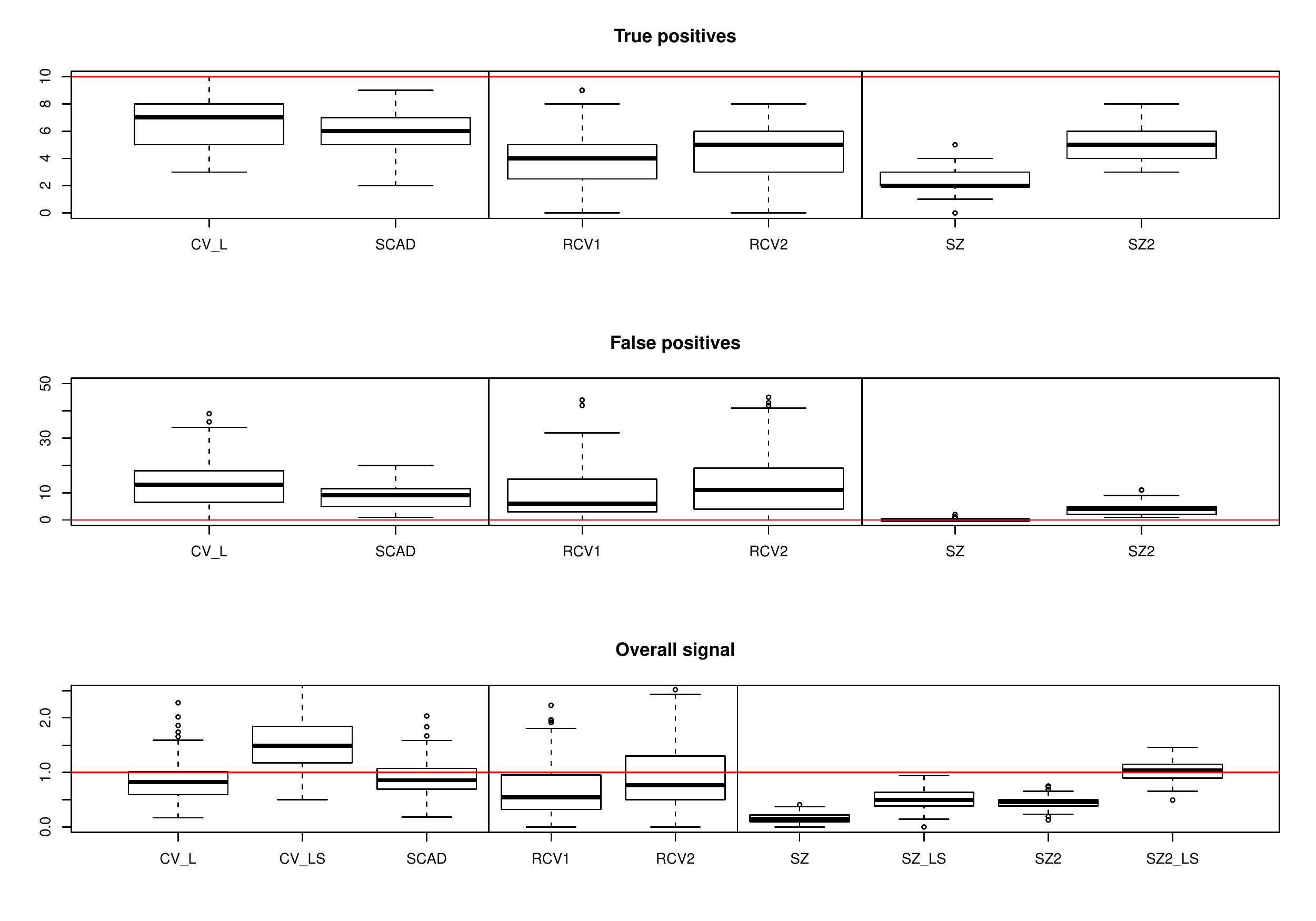}
\caption{\emph{Diagnostic plot. Top panel shows boxplots of the number of non-zero coefficients correctly identified by each procedure over the $B = 100$ simulation runs. Middle panel shows the number of zero coefficients incorrectly given non-zero values. Bottom panel shows the ratio of the estimated signal to the true signal $\frac{\sum_{j = 1}^p|\hat{\beta}_j|}{\sum_{j = 1}^p|\beta_j^*|}$. $p = 100$, $\alpha = 0.5$ (so that there are 10 signal variables and 90 zero variables) and $snr = 1$.}}
\label{figDiagAlpha0.5SNR1p100}
\end{figure}

Figure~\ref{figDiagAlpha0.5SNR1p100} is a diagnostic plot showing three measures pertaining to the quality of the estimated $\beta$ for each of the methods (CV Lasso, SCAD, SZ, SZ2 and both halves of the RCV - labelled RCV1 and RCV2). Parameters for this figure are $p = 100$, $\alpha = 0.5$ and $snr = 1$. This is one of many such figures that can be drawn, but this one is representative and is all that is presented, in interest of saving space.

Notice that both CV\_L and SCAD tend to select more of the signal variables than do the other variables. None of the methods select all the signal variables. This would lead to considerable upward bias as signal size increases, as the residual sum of squares on which all of these estimators are based would inflate with the signal not successfully removed from it.

CV\_L and SCAD seem to counter this shortcoming by selecting a moderate number of irrelevant variables  and giving them non-zero coefficients (middle). The balance between missing signal variables and capturing irrelevant variables seems to lead to an estimated coefficient vector with signal size rather close to that of the true parameter vector (bottom panel).

RCV seems to select too few signal variables, making it difficult to strike a balance to find a decent variance estimate. SZ and SZ2 both select fewer true signal variables than CV\_L and SCAD and detect almost no false positives. The signal variables not selected by the SZ and SZ2 estimators then degrade their performance as signal size increases. Notice that neither RCV, SZ nor SZ2 produces signal sizes large enough to match the underlying signal (bottom panel). Least squares estimates tend to have signals larger than the true signal, because they have the same set of nonzero coefficients as their penalised counterparts, but with larger, unpenalised coefficients.

Large upward biases occur because none of the methods select all the true signal variables. Some methods seem to find a balance between selecting signal and non-signal variables to produce reasonably good variance estimates over a broad range of sparsity and signal size settings. Quite why CV\_L and SCAD behave this way is not fully understood, but we suspect that the adaptive selection of the regularisation parameter contributes.

\subsubsection{Ranging over different $\alpha$}

\begin{figure}[htb]
  \centering
  \includegraphics[width=140mm, height=100mm]{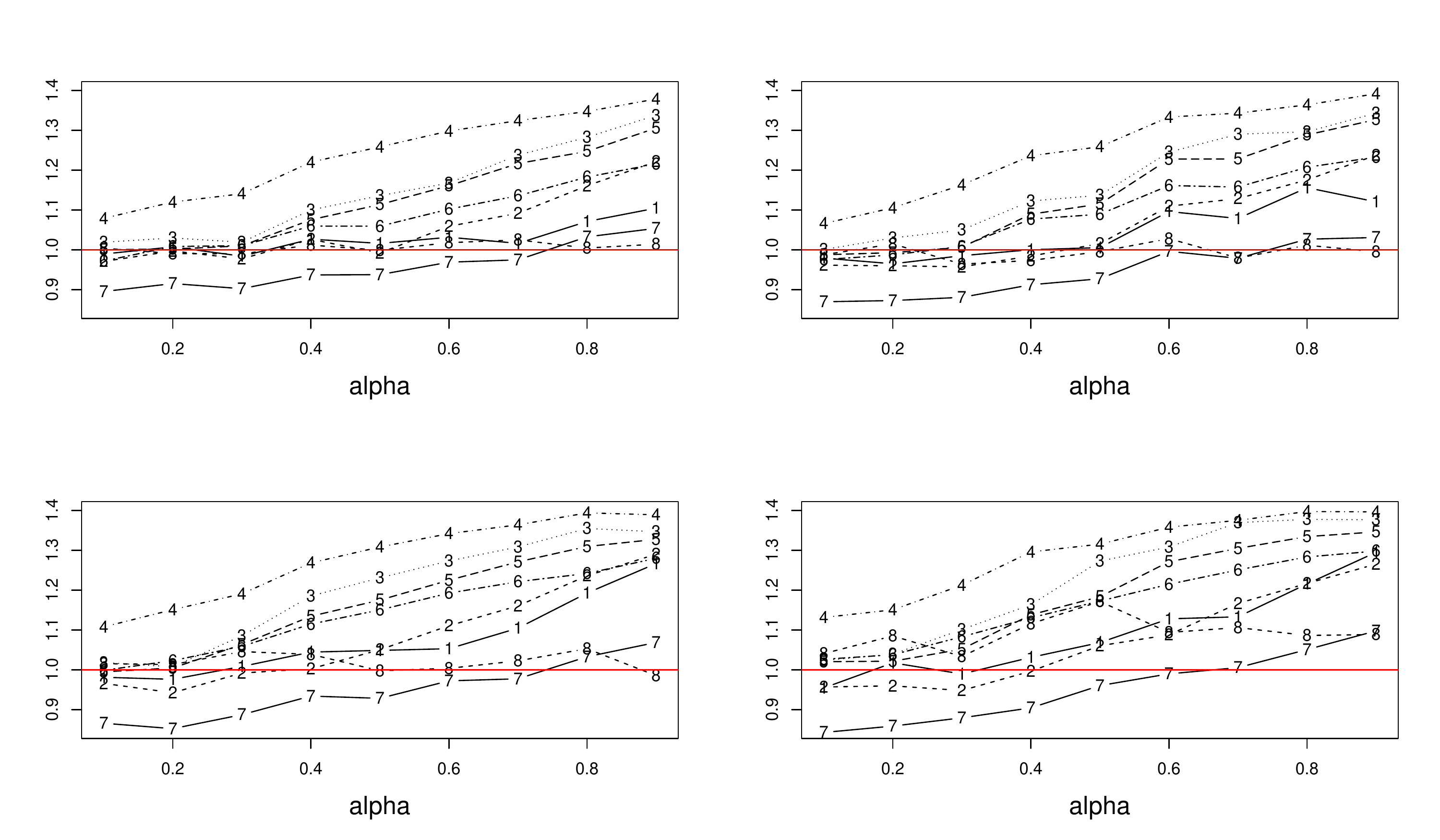}
  \caption{\emph{Median standard deviation estimates over different levels of $\beta$ sparsity. Plot numbers refer to CV\_L (1), SCAD (2), RCV (3), SZ (4), SZ\_LS (5), SZ2 (6), SZ2\_LS (7) and D1 (8) respectively. $\sigma = 1$}}
  \label{alphaPlotRho0SNR1}
\end{figure}

Figure~\ref{alphaPlotRho0SNR1} plots median standard deviation estimates over different values of $\alpha$. Here we set $snr = 1$ and $\sigma = 1$. Notice how CV\_L and SCAD resist upward bias over a broader range of sparsity settings, with CV\_L performing most admirably for smaller values of $p$ (top row). This is revealed in the figure by lines 1 and 2 hugging the red reference line (true $\sigma$) quite closely for $\alpha$ up to 0.5, while by this time, all the other curves have diverged significantly.

It is interesting to note that the median method of moments estimators are largely immune to a decrease in sparsity (increase in $\alpha$). Remember that this comes at the expense of larger estimator variance, as reflected in the preceding figures. 

\subsection{Effect of signal-to-noise ratio}
There are two components contributing to the size of $\beta$: the degree of sparsity and the per element signal size. For a given sparsity level (number of nonzero elements of $\beta$), the higher the SNR (as defined earlier), the higher the per element signal strength. We found in our simulations that individual signal sizes have significant impact on the quality of variance estimates. 

\begin{figure}[htb]
  \centering
  \includegraphics[width=140mm, height=100mm]{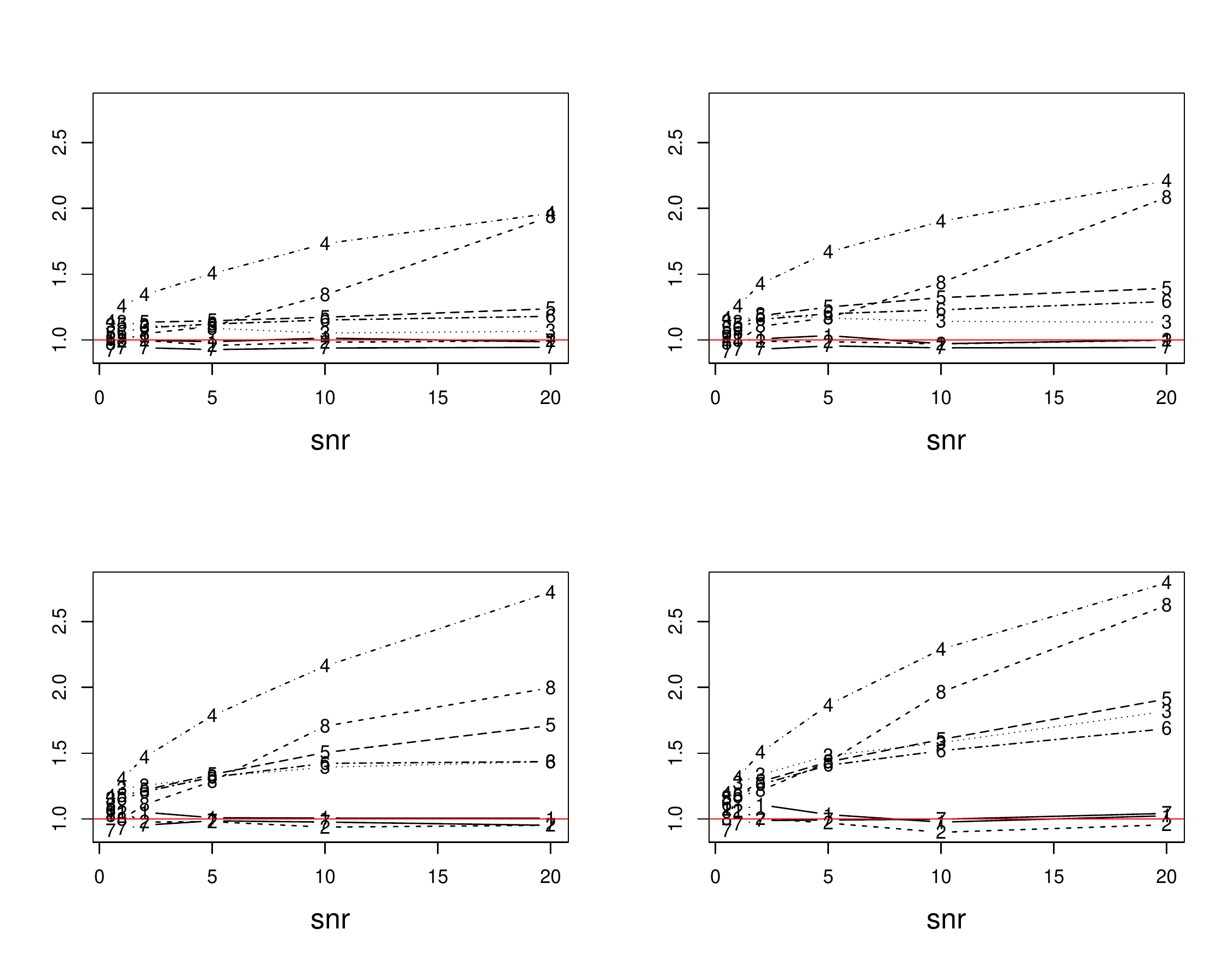}
  \caption{\emph{Median standard deviation estimates over different levels of signal-to-noise ratio. Plot numbers refer to CV\_L (1), SCAD (2), RCV (3), SZ (4), SZ\_LS (5), SZ2 (6), SZ2\_LS (7) and D1 (8) respectively. $\alpha = 0.5$, $\sigma = 1$.}}
  \label{snrPlotRho0alpha0.5}
\end{figure}

Figure~\ref{snrPlotRho0alpha0.5} is a telling demonstration of the superiority of the CV\_L and SCAD estimators (i.e. those with data dependent, adaptively selected regularisation parameters). Sparsity level is set at $\alpha = 0.5$, a level both theoretically and anecdotally significant. Theoretical results suggest that, at this level of sparsity, all estimators considered are consistent. This asymptotic result is falsely comforting in finite samples. Clearly some of the estimators are significantly upwardly biased when the signal strength increases.

Anecdotally, it seems as though this level of sparsity coincides with a point of deterioriation of our estimators. As the $\beta$ vector becomes less sparse beyond this point, the performance of all estimators deteriorates rapidly, suggesting that this level is a significant watershed beyond which we have little hope of decent error variance estimates. As we skirt this precarious edge by increasing the per element signal, we see that CV\_L and SCAD remain unaffected, while all other candidates suffer significantly. Although not shown, these plots look similar for the high noise ($\sigma = 3$) case.

Interestingly, the least-squares-after-scaled-Lasso estimator with the smaller regularisation parameter (SZ2\_LS) seems to perform admirably here as well. This, however, is an artefact of setting the sparsity level at $\alpha = 0.5$. For all other sparsity levels, this estimator exhibits significant biases in either direction.

\subsection{Effect of predictor correlation: changing $\rho$}
\begin{figure}[htb]
  \centering
  \includegraphics[width=140mm, height=100mm]{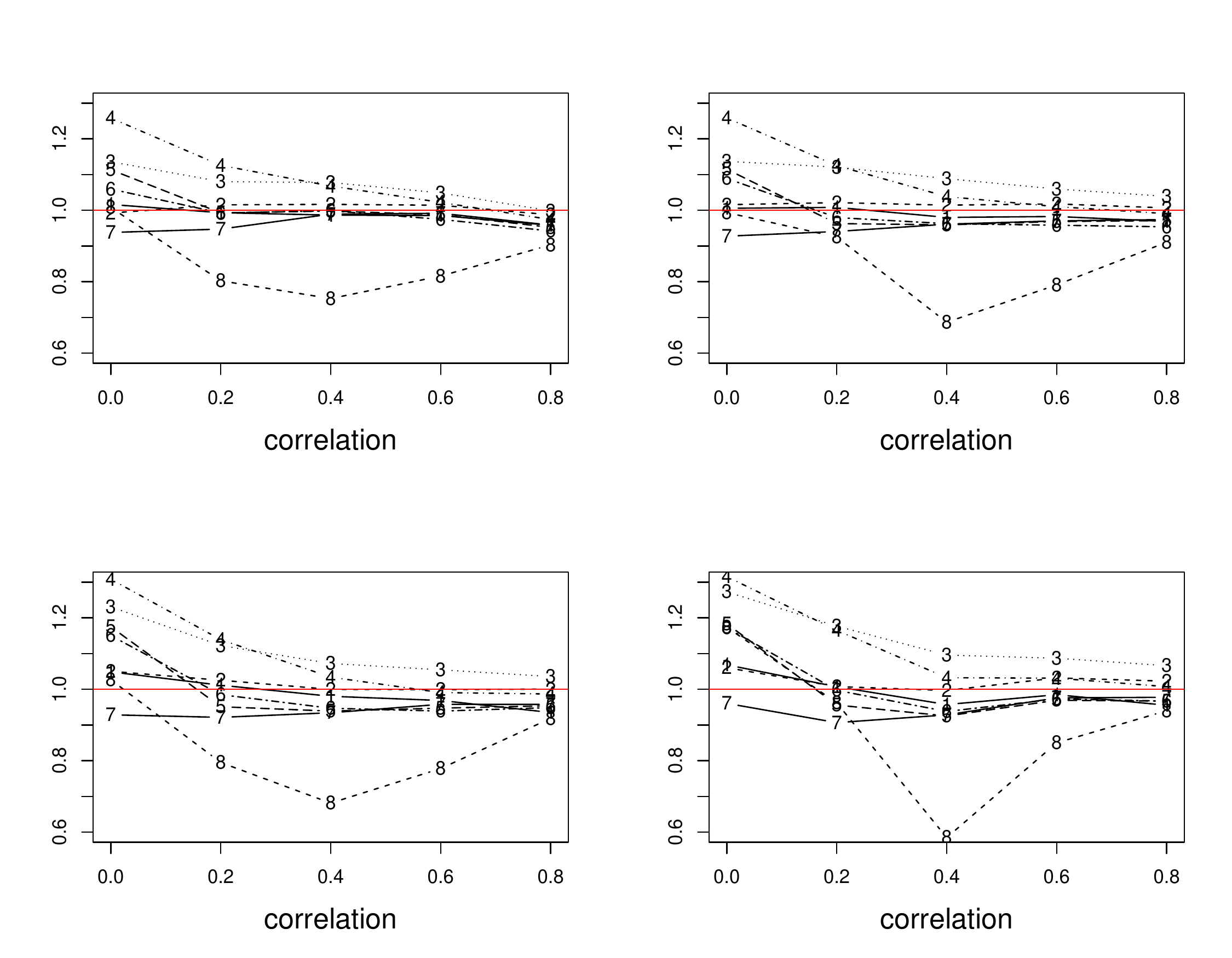}
  \caption{\emph{Median standard deviation estimates over different levels of predictor correlation. Sample size $n = 100$ and predictor numbers $p = 100, 200, 500, 1000$ left to right over rows. Plot numbers refer to CV\_L (1), SCAD (2), RCV (3), SZ (4), SZ\_LS (5), SZ2 (6), SZ2\_LS (7) and D2 (8) respectively. $\alpha = 0.5$, $\sigma = 1$, $snr = 1$.}}
  \label{rhoPlotAlpha0.5}
\end{figure}

It is interesting to note that correlation between predictors seems to come to the rescue of some of the variance estimators here considered. Figure~\ref{rhoPlotAlpha0.5} again plots median standard deviations, this time as a function of predictor correlation ($\rho$). Notice how the large upward bias of the RCV, SZ and SZ2 estimators decreases as $\rho$ increases. Unfortunately, the method of moments estimators perform rather poorly as predictor correlation increases, even D2, which is supposed to be designed for general predictor correlation structures.

\section{Orthogonal predictor matrix and a certainty equivalent variance estimator}
Obtaining finite sample results (or even asymptotic results) about variance estimators with adaptively chosen regularisation parameters seems like a difficult task. In this section, we consider a very simple setup which allows for some tractable results. In particular, since most error variance estimators are based on residual sum of squares, we wish to study the behaviour of a variance estimator based on this quantity in a simple scenario.

Consider the orthogonal case where $p = n$ and $X = I_n$, the $n \times n$ identity matrix. In this case, we have each $Y_i \sim N(\beta_i, \sigma^2)$. We assume that sparsity of the $\beta$ vector is governed by $\alpha < 1$. In particular, we have $\beta_i = \beta$ for $i = 1, 2, ..., \lceil n^\alpha \rceil$ and $\beta_i = 0$ otherwise. Call this the \textit{orthogonal sparsity model}. 

Estimates of $\beta_i$ are obtained by minimising the objective
\[
\sum_{i = 1}^n(Y_i - \beta_i)^2 + \lambda\sum_{i = 1}^n|\beta_i|
\]
to obtain the solution $\hat{\beta}_i = S(Y_i, \lambda)$, where $S(x, t) = {\rm{sign}}(x)\max\{|x|-t, 0\}$ is the soft thresholding operator with threshold $t$. Plugging these quantities into the residual sum of squares, we obtain:
\begin{align*}
  RSS &= \sum_{i = 1}^n(Y_i - \hat{\beta}_i)^2\\
 &= \sum_{i = 1}^n\min\{Y_i^2, \lambda^2\}
\end{align*}
from which we derive the family of estimators for $\sigma^2$ (indexed by $\lambda$):
\begin{equation}
  \hat{\sigma}^2_{n, \lambda} = \frac{\sum_{i = 1}^n \min\{Y_i^2, \lambda^2\}}{\sum_{i = 1}^n I\{|Y_i| \leq \lambda\}}
\label{sigEstiOrthog}
\end{equation}

To make this a practicable estimate of $\sigma^2$, we need to select a single member from the family (i.e. a value for $\lambda$). Many are possible, but in light of the discussion of previous sections, let us select it adaptively. Consider then a single held out set $Z_i \stackrel{d}{=} Y_i$ with corresponding cross-validation error
\[
CV(\lambda) = \sum_{i = 1}^n(Z_i - S(Y_i, \lambda))^2
\]

The adatively chosen regularisation parameter is then:
\begin{equation}
  \tilde{\lambda} = {\rm argmin}_\lambda CV(\lambda)
 \label{cvRegParam}
\end{equation}

Although an interesting prospect - and one in keeping with the arguments of the paper - its theoretical properties are not considered here. We believe that the estimator $\hat{\sigma}^2_{n, \tilde{\lambda}}$ is amenable to theoretical analysis and that such an analysis may be instructive to the workings of variance estimators with adaptively chosen regularisation parameters. This is definitely a channel for (immediate) future investigation. However, in the sequel we consider the behaviour of estimators \eqref{sigEstiOrthog} under deterministic sequences $\lambda_n$. After some general results, we consider a specific sequence, dubbed the \textit{certainty equivalent} (CE) sequence of $\lambda$ and denoted $\hat{\lambda}_n$, which bears resemblance to the adaptive selection \eqref{cvRegParam}. Finally, a small simulation reveals how similarly $\hat{\sigma}^2_{n, \tilde{\lambda}}$ and $\hat{\sigma}^2_{n, \hat{\lambda}_n}$ behave in small samples, giving hope that the results gleaned for the later apply to the former.

\subsection{General deterministic sequences: $\lambda_n$}
Our first result considers the large sample behaviour of the denominator in \eqref{sigEstiOrthog} under a determinsitic sequence $\lambda_n$:
\begin{lemma}
If $\lambda_n \rightarrow \infty$ as $n \rightarrow \infty$, then under the orthogonal sparsity model, $\frac{1}{n}\sum_{i = 1}^n I\{|Y_i| > \lambda_n\} \stackrel{P}{\rightarrow} 0$.
\label{lemmaDenomDeter}
\end{lemma}

All theorems and lemmas are proved in the appendix. Lemma \ref{lemmaDenomDeter} suggests that we need not consider $\hat{\sigma}^2_{n, \lambda_n}$ directly, but rather the more tractable
\[
\tilde{\sigma}^2_{n, \lambda_n} = \frac{1}{n}\sum_{i = 1}^n\min\{Y_i^2, \lambda^2_n\}
\]

The next lemma characterises the limiting expectation and variance of $\tilde{\sigma}^2_{n, \lambda_n}$. This is an intermediate result from which the consistency of $\hat{\sigma}^2_{n, \lambda_n}$ for $\sigma^2$ follows almost immediately.
\begin{lemma}
If $\lambda_n \rightarrow \infty$ as $n \rightarrow \infty$, then under the orthogonal sparsity model
\[
 E[\tilde{\sigma}^2_{n, \lambda_n}] \rightarrow \sigma^2
\]
\[
Var[\tilde{\sigma}^2_{n, \lambda_n}] \rightarrow 0
\]
\label{lemmaExpVarDeter}
\end{lemma}

Putting these two lemmas together, it is not too difficult to prove the following theorem:
\begin{theorem}
  if $\lambda_n \rightarrow \infty$ as $n \rightarrow \infty$, then under the orthogonal sparsity model
\[
\hat{\sigma}^2_{n, \lambda_n} \stackrel{P}{\rightarrow} \sigma^2
\]
\[
\sqrt{n}(\hat{\sigma}^2_{n, \lambda_n} - \sigma^2) \stackrel{d}{\rightarrow} N(0, 2\sigma^4)
\]
  \label{theoConsisANDeter}
\end{theorem}

Note that consistency and asymptotic normality (with asymptotic variance equal to that of the oracle estimator) are not too hard to come by in this family of estimators. All we need to is select $\lambda_n$ that tends to $\infty$ with $n$. This is somewhat surprising, but meshes nicely with evidence from the simulation studies earlier in the paper. An estimator can have these desirable asymptotic properties, but since the requirement on $\lambda_n$ to achieve these properties is weak, many consistent, asymptotically normal estimators can have poor finite sample performance.

For example, suppose we set $\lambda_n = \infty$, so that $\hat{\sigma}^2_{n, \lambda_n} = \frac{1}{n}\sum_{i = 1}^n Y_i$. This estimator satisfies the conditions of Theorem \ref{theoConsisANDeter}, but has finite sample expectation $\sigma^2 + \frac{n^\alpha}{n}\beta^2$. Notice how we can make this estimator arbitrarily biased upward in a finite sample by merely increasing the signal strength $\beta$ or reducing sparsity (increasing $\alpha$). We quest, then, for a $\lambda_n \rightarrow \infty$, but chosen so as to have good finite sample performance as well. 

\subsection{Certainty equivalent sequence $\hat{\lambda}_n$}

Instead of choosing $\lambda$ as the (random) minimiser of $CV(\lambda)$ and inducing dependence between the summands of the numerator of $\hat{\sigma}^2_{n, \tilde{\lambda}}$ (further increasing complexity), we could choose a deterministic sequence (hopefully) bearing close relation to it. In particular, we can minimise $ECV_n(\lambda, \beta, \alpha) = E[CV(\lambda)]$, which can be written as:
\[
ECV_n(\lambda, \beta, \alpha) =  n\sigma^2 + \frac{n^\alpha}{n} \cdot r_S(\lambda, \beta) + \frac{n-n^\alpha}{n}\cdot r_S(\lambda, 0)
\]
where $r_S(\lambda, \beta) = E(S(Y_i, \lambda) - \beta)^2$ is the risk of the soft thresholding operator, for which we have the expression (\citet{JohnstoneGE}):
\begin{align*}
 r_S(\lambda, \beta) &= \sigma^2 + \lambda^2 + (\beta^2 - \lambda^2 - \sigma^2)\left[\Phi\left(\frac{\lambda-\beta}{\sigma}\right) - \Phi\left(\frac{\lambda-\beta}{\sigma}\right)\right] \\
&- \sigma(\lambda - \beta)\phi(\lambda + \beta) - \sigma(\lambda + \beta)\phi(\lambda - \beta)
\end{align*}
The ``certainty equivalent'' choice for $\lambda$ then becomes:
\[
\hat{\lambda}_n = \hat{\lambda}_n(\beta, \alpha) = {\rm{argmin}}_{\lambda}R_n(\lambda, \beta, \alpha)
\]
where $R_n(\lambda, \beta, \alpha) = \frac{n^\alpha}{n} r_S(\lambda, \beta) + \frac{n-n^\alpha}{n}r_S(\lambda, 0)$.

\begin{figure}[htb]
 \centering
 \includegraphics[width=140mm, height=100mm]{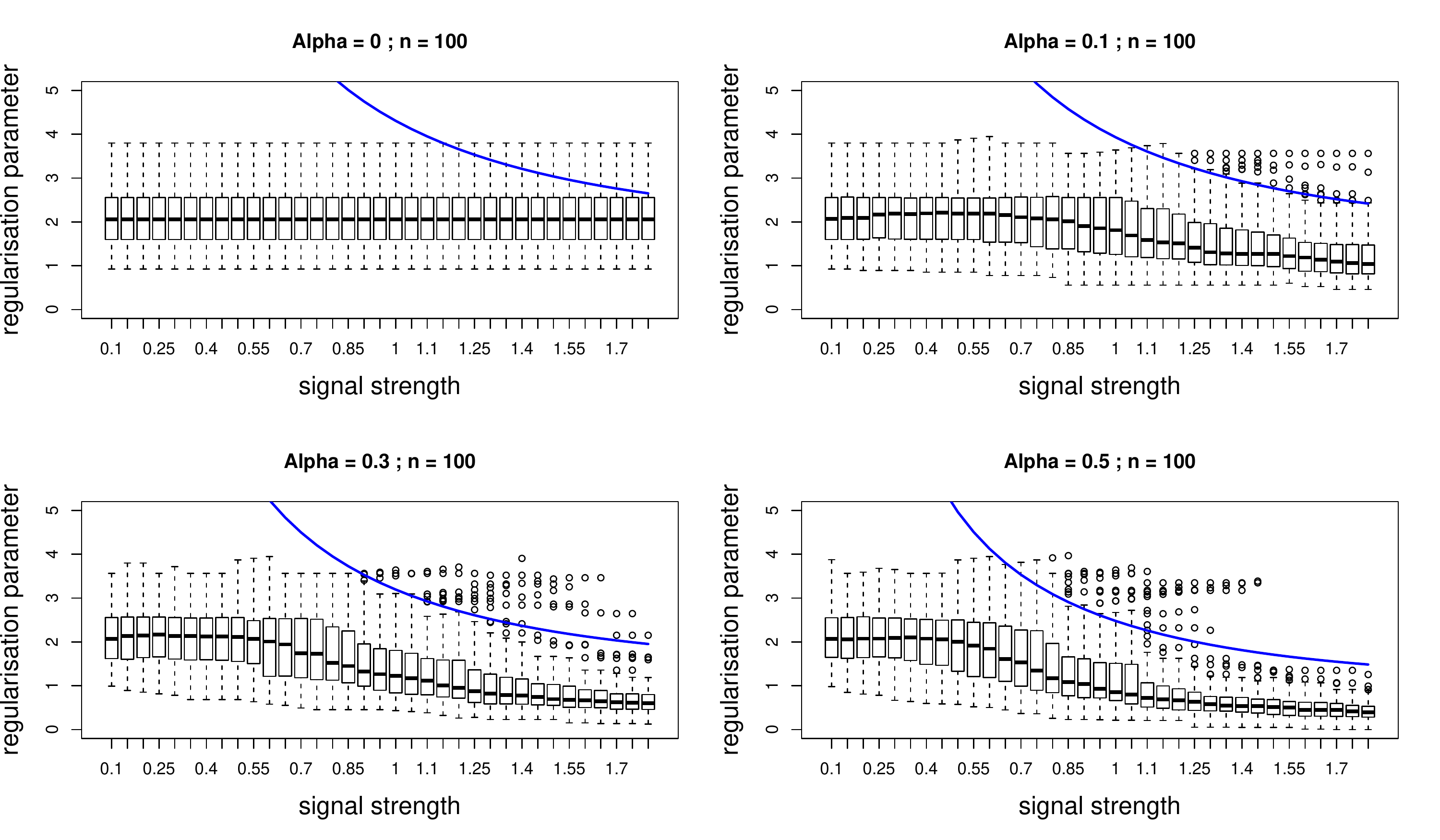}
 \caption{\emph{Certainty equivalent regularisation parameter as a function of signal strength ($\beta$), at different sparsity levels ($\alpha$) along with reference boxplots of $CV(\lambda)$ minimising regularisation parameter. Top left panel corresponds to $\alpha = 0$; top right, $\alpha = 0.1$; bottom left, $\alpha = 0.3$ and bottom right, $\alpha = 0.5$. A sample size of $n = 100$ is used when generating replications of CV minimising $\tilde{\lambda}$.}}
 \label{figRegParam}
\end{figure}

The optimisation can be done numerically. Figure~\ref{figRegParam} plots $\hat{\lambda}_n(\beta, \alpha)$ as a function of the signal strength $\beta$ for four different levels of sparsity $\alpha$ (red curves). Also plotted in Figure~\ref{figRegParam} are boxplots gleaned from $B = 100$ realisations of $\tilde{\lambda}$ where
\[
\tilde{\lambda} = {\rm{argmin}}_\lambda CV(\lambda)
\]
for a sample of size $n = 100$. We plot these for reference, because they are the realisations of the random regularisation parameter we would actually compute in an application. Notice how $\hat{\lambda}_n$ tends to lie everywhere above the rump of the $\tilde{\lambda}$ values at a given signal strength, exhibiting a similar shape. Although convenient theoretically, the certainty equivalent estimate of $\lambda$ does not seem to accord with the random estimate obtained by minimising $CV(\lambda)$. Despite this, their estimates for $\sigma^2$ are not too different, as demonstrated in the next section. 

Note that this sequence cannot be obtained in real applications, because we do not know $\sigma^2$ nor $\beta$. The certainty equivalent sequence can only be generated by an oracle. The true utility of this sequence comes from its theoretical tractability. As a function of $\lambda$, $r_S(\lambda, 0)$ is monotonically decreasing, convex and non-negative. Its minimum value is $0$, achieved by setting $\lambda = \infty$. For $\beta \neq 0$, $r_S(\lambda, \beta)$ is initially decreasing in $\lambda$, attains a unique global minimum, whereafter is is non-decreasing in $\lambda$, with horizontal asymptote $\beta^2$. Similar properties are shared by $\frac{n^\alpha}{n}r_S(\beta, \lambda)+\frac{n-n^\alpha}{n}r_S(\lambda, 0)$. 

We can show that $\hat{\lambda}_n \rightarrow \infty$, making $\hat{\sigma}^2_{n, \hat{\lambda}_n}$ consistent for $\sigma^2$ and ensuring it has an asymptotic normal distribution (from the previous section). Furthermore, one can show how this sequence minimises an upper bound to the upward bias of the estimator in small samples. Downward bias does not seem to be a problem for this estimator (see Figure \ref{figScaledRSS}).

\begin{theorem}
  Assume the orthogonal sparsity model. The certainty equivalent sequence $\hat{\lambda}_n \rightarrow \infty$ as $n \rightarrow \infty$.
  \label{theoCEInf}
\end{theorem}

To see how this sequence minimises an upper bound on the upward bias, consider the Stein Unbiased Risk Estimate (SURE) for $r_S(\lambda, \beta)$ (\citet{JohnstoneGE}):
\[
r_S(\lambda, \beta) = E[\sigma^2 - 2\sigma^2I\{|Y| \leq \lambda\} + \min\{Y^2, \lambda^2\}] 
\]
where $Y \sim N(\beta, \sigma^2)$, so that
\begin{align*}
  E[\min\{Y_i^2, \lambda\}] - \sigma^2 &= r_S(\lambda, \beta_i) - 2P(|Y_i| > \lambda)\\
&\leq r_S(\lambda, \beta_i)
\end{align*}
which translates to
\[
E[\tilde{\sigma}^2_{n, \lambda_n}]- \sigma^2 \leq \frac{n^\alpha}{n}r_S(\lambda_n, \beta) + \frac{n-n^\alpha}{n}r_S(\lambda_n, 0)
\]
the right hand side of which is minimised by the certainty equivalent sequence of $\lambda$. The certainty equivalent sequence minimises an upper bound to the bias, making a concerted effort to keep it as small as possible \textit{in the finite sample}.

\subsection{Comparison of CE and CV variance estimators}

\begin{figure}[htb]
 \centering
 \includegraphics[width=140mm, height=100mm]{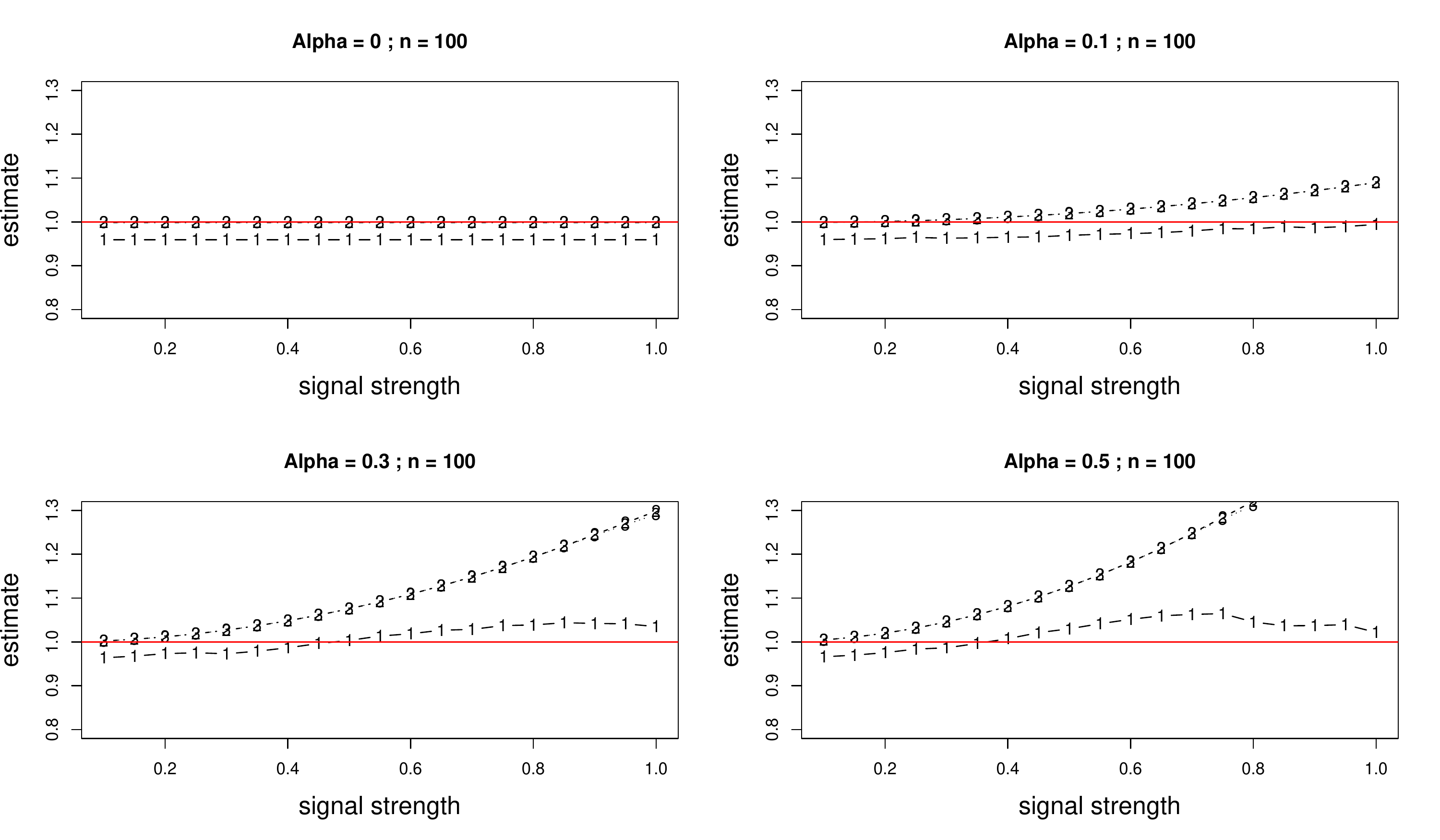}
 \caption{\emph{Variance estimates as a function of signal strength ($\beta$), at different sparsity levels ($\alpha$). Curves labelled ``1'' plot the CV estimate, while those labelled ``2'' and ``3'' the CE estimates (with different denominators). Sparsity parameter $\alpha$ = 0, 0.1, 0.3 and 0.5, varying left to right along rows and then down along columns. Red horizontal lines show the true variance.}}
 \label{figScaledRSS}
\end{figure}

Figure~\ref{figScaledRSS} plots, for different levels of sparsity in different panels, the mean over $B = 100$ replications of three variance estimators. The curves labelled ``1'' are the means of the replications of $\hat{\sigma}^2_{CV} = \hat{\sigma}^2_{n, \tilde{\lambda}}$, where $\tilde{\lambda}$ is chosen according to Equation~\eqref{cvRegParam}. Curves labelled ``2'' are of the means of the replications of $\hat{\sigma}^2_{CE} = \hat{\sigma}^2_{n, \hat{\lambda}_n}$, where $\hat{\lambda}_n$ is the certainty equivalent sequence of $\lambda$, while those labelled ``3'' are of $\tilde{\sigma}^2_{CE} = \tilde{\sigma}^2_{n, \hat{\lambda}_n}$. Notice how close curves ``2'' and ``3'' are to each other. We have the theoretical guarantee on the bias of curves ``3''.

Note that the three estimators behave reasonably similarly, except for low sparsity, high signal cases. It is heartening to note that the CE estimate seems to suffer from upward bias in this case, despite its guarantee of a minimum upper bound on this bias. The CV estimator actually seems to do an even better job of selecting the appropriate regularisation parameter in the small sample setting - a clear case for further analysis of its properties.

\section{Discussion}
Error variance estimation in linear regression when $p > n$ is a difficult problem that deserves attention. Several estimators have been proposed. We have reviewed these and some of the theoretical results around them. Despite some comforting asymptotic results, finite sample performance of these estimators seems to suffer, particularly when signals become large and non sparse. 

Variance estimators based on residual sums of squares with adaptively chosen regularisation parameters seem to have promising finite sample properties. In particular, we recommend the cross-validation based, Lasso residual sum of squares estimator as a good variance estimator under a broad range of sparsity and signal strength assumptions. The complexity of their structure seems to have discouraged their rigorous analysis. Simulation results from this paper seem to suggest that there could be value in understanding these estimators more fully.

\section{Acknowledgements}
We wish to thank Iain Johnstone for his helpful comments that seeded the analysis of the certainty equivalent in the orthogonal case.

\bibliographystyle{agsm}
\bibliography{tibs}

\section*{Appendix - Proofs of lemmas and theorems}
\subsection*{Proof of Lemma \ref{lemmaDenomDeter}}
Assume the orthogonal sparsity model and consider
\begin{align*}
  E&\left[\frac{1}{n}\sum_{i = 1}^nI\{|Y_i| > \lambda_n\}\right] \\ &= \frac{1}{n}\sum_{i = 1}^n P(|Y_i| > \lambda_n) \\
&= \frac{n^\alpha}{n}\left(1 - \Phi\left(\frac{\lambda_n-\beta}{\sigma}\right) + \Phi\left(\frac{-\lambda_n-\beta}{\sigma}\right)\right) + 2\frac{n-n^\alpha}{n}\left(1-\Phi\left(\frac{\lambda_n}{\sigma}\right)\right)\\
&\leq 2\frac{n^\alpha}{n} + 2\frac{n-n^\alpha}{n}\left(1-\Phi\left(\frac{\lambda_n}{\sigma}\right)\right) \\
&\rightarrow 0
\end{align*}
as $n, \lambda_n \rightarrow \infty$. The result follows upon the application of Markov's inequality.

\subsection*{Proof of Lemma \ref{lemmaExpVarDeter}}
Assume the orthogonal sparsity model and $\lambda_n \rightarrow \infty$ as $n \rightarrow \infty$ and consider:
\begin{align*}
E&\left[\min\{Y_i^2, \lambda_n^2\}\right] \\
&= E[Y_i^2; -\lambda_n \leq Y_i \leq \lambda_n] + \lambda_n^2P(|Y_i| > \lambda_n)\\
&= E[Y_i^2; -\lambda_n \leq Y_i \leq \lambda_n] + \lambda_n^2\left(1-\Phi\left(\frac{\lambda_n - \beta_i}{\sigma}\right)\right) + \lambda_n^2\left(1-\Phi\left(\frac{\lambda_n+\beta_i}{\sigma}\right)\right)
\end{align*}
Now
\[
E[Y_i^2; -\lambda_n \leq Y_i \leq \lambda_n] = \sigma^2m_2(\lambda_n, \beta_i) + 2\sigma\beta_im_1(\lambda_n, \beta_i) + \beta_i^2m_0(\lambda_n, \beta_i)
\]
where
\[
m_j(\lambda, \beta) = \int_{\frac{-\lambda-\beta}{\sigma}}^{\frac{\lambda - \beta}{\sigma}}x^j\phi(x)\,dx
\]
for all non-neative integers $j$. Note that for fixed $\beta$ and $\sigma$ and $\lambda \rightarrow \infty$, these tend to the $j^{th}$ moments of the standard normal distribution. So $m_0(\lambda_n, \beta_i) \rightarrow 1$, $m_1(\lambda_n, \beta_i) \rightarrow 0$ and $m_2(\lambda_n, \beta_i) \rightarrow 1$ as $n \rightarrow \infty$. This suggests that $E[Y_i^2; -\lambda_n \leq Y_i \leq \lambda_n] \rightarrow \sigma^2 + \beta_i^2$ as $n \rightarrow \infty$. 

Also note that for $x > 0$, as $x \rightarrow \infty$, $x^k\left(1-\Phi(x)\right) \sim x^{k-1}\phi(x) \rightarrow 0$ for any finite integer $k > 1$. Hence
\begin{align*}
E[\tilde{\sigma}^2_{n, \lambda_n}] &\sim \frac{n^\alpha}{n}(\sigma^2 + \beta^2) + \frac{n - n^\alpha}{n}\sigma^2 \\
&\rightarrow \sigma^2
\end{align*}
as $n \rightarrow \infty$.

Similarly,
\begin{align*}
  E&[\min\{Y_i^4, \lambda^4\}] \\
 &= E[Y_i^4; -\lambda_n \leq Y_i \leq \lambda_n] + \lambda_n^4\left(1-\Phi\left(\frac{\lambda_n - \beta_i}{\sigma}\right)\right) + \lambda_n^4\left(1-\Phi\left(\frac{\lambda_n+\beta_i}{\sigma}\right)\right)
\end{align*}
with
\begin{align*}
E[Y_i^4; -\lambda_n \leq Y_i \leq \lambda_n] &= \sigma^4m_4(\lambda_n, \beta_i) + 4\sigma^3\beta_im_3(\lambda_n, \beta_i) + 6\sigma^2\beta_i^2m_2(\lambda_n, \beta_i)\\ 
&+ 4\sigma\beta_i^3m_1(\lambda_n, \beta_i) + \beta_i^4m_0(\lambda_n, \beta_i) \\
&\sim 3\sigma^4 + 6\sigma^2\beta_i^2 + \beta_i^4
\end{align*}
Also
\begin{align*}
E[\min\{Y_i^2, \lambda_n^2\}\min\{Y_j^2, \lambda_n^2\}] &= E[\min\{Y_i^2, \lambda_n^2\}]E[\min\{Y_j^2, \lambda_n^2\}] \\
 &\sim \sigma^4 + \sigma^2\beta_i^2 + \sigma^2\beta_j^2 + \beta_i^2\beta_j^2
\end{align*}
So that
\begin{align*}
  E[\tilde{\sigma}^4_{n, \lambda_n}] &\sim \frac{n^\alpha}{n^2}(3\sigma^4 + 6\sigma^2\beta^2 + \beta^4) + \frac{n-n^\alpha}{n^2}(3\sigma^4) + \frac{n^\alpha(n^\alpha -1)}{n^2}(\sigma^4 + 2\sigma^2\beta^2 + \beta^4) \\
&+ \frac{n^\alpha(n-n^\alpha)}{n^2}(\sigma^4 + \sigma^2\beta^2) + \frac{(n-n^\alpha)(n-n^\alpha-1)}{n^2}(\sigma^4)\\
&\rightarrow \sigma^4
\end{align*}
as $n \rightarrow \infty$. Hence $Var[\tilde{\sigma}^2_{n, \lambda_n}] \rightarrow 0$.

\subsection*{Proof of Theorem \ref{theoConsisANDeter}}
An application of Markov's inequality to $(\tilde{\sigma}^2_{n, \lambda_n} - \sigma^2)^2$, combined with the results of Lemma \ref{lemmaExpVarDeter} give us consistency of $\tilde{\sigma}^2_{n, \lambda_n}$ for $\sigma^2$. Combined with the result of Lemma \ref{lemmaDenomDeter}, we have that $\hat{\sigma}^2_{n, \lambda_n}$ is consistent for $\sigma^2$.

Asymptotic normality of $\tilde{\sigma}^2_{n, \lambda_n}$ follows from a central limit theorem applied to the \textit{independent} summands of the numerator. The asymptotic variance is gleaned from the proof of Lemma \ref{lemmaExpVarDeter}, by noting that the results quoted there imply that $nVar[\tilde{\sigma}^4_{n, \lambda_n}] \sim 3\sigma^4 - \sigma^4 = 2\sigma^4$. Lemma \ref{lemmaDenomDeter} ensures that this asymptotic normality holds for $\hat{\sigma}^2_{n, \lambda}$ as well.

\subsection*{Proof of Theorem \ref{theoCEInf}}
Fix $\alpha$ and $\beta$ and let $f_n(\lambda) = R_n(\lambda, \beta, \alpha)$ and $g(\lambda) = r_S(\lambda, 0)$. Note that $f_n(\lambda) \rightarrow g(\lambda)$ for all $\lambda$ as $n \rightarrow \infty$.
Let $\lambda_- = {\rm liminf} \hat{\lambda}_n$. Assume $\lambda_- < \infty$. Now $f_n(\lambda_-)$ converges to $g(\lambda_-)$ and $f_n(\infty)$ converges to $g(\infty) = 0$. Since $g(\lambda_-) > 0$, $f_n(\lambda_-) \leq f_n(\infty)$ only finitely many times. 

However, when $\hat{\lambda}_n\leq \lambda_-$, $f_n(\lambda_-) \leq f_n(\infty)$, because these functions are increasing to the right of $\lambda_-$. This occurs infinitely many times, which is a contradiction. Hence $\lambda_- = {\rm liminf} \hat{\lambda}_n = \infty$ and ${\rm lim} \hat{\lambda}_n = \infty$.

\end{document}